\documentclass[pra, superscriptaddress, aps, 10pt, twocolumn, showkeys, a4paper]{revtex4-2}
\usepackage{amssymb}
\usepackage{amsmath}
\usepackage{graphicx}
\usepackage{xcolor}
\usepackage{tabularx}
\usepackage{hyperref}
\usepackage{float}

\begin{document}

\title{Strengthening the No-Go Theorem for QRNGs}

\author{Vardaan Mongia} \email{vardaan.mongia@gmail.com}
\affiliation{Quantum Science and Technology Laboratory, Physical Research Laboratory, Ahmedabad, 380009, India}
\affiliation{Dept. of Physics, Indian Institute of Technology Gandhinagar, Gujarat, 382355, India}

\author{Abhishek Kumar}
\affiliation{Space Weather Laboratory, Physical Research Laboratory, Ahmedabad, 380009, India}

\author{Shashi Prabhakar} \email{shaship@prl.res.in}
\affiliation{Quantum Science and Technology Laboratory, Physical Research Laboratory, Ahmedabad, 380009, India}

\author{R. P. Singh}
\affiliation{Quantum Science and Technology Laboratory, Physical Research Laboratory, Ahmedabad, 380009, India}

\date{\today} 

\begin{abstract}
Quantum random numbers are essential for security against quantum algorithms. Randomness as a beacon is a service being provided for companies and governments to upgrade their security standards from RSA to PQC-QKD or PQC-RSA protocols. Both security mechanisms assume trust in the service provider unless one aims for device-independent protocols. How does an entity ensure that the beacon service has a quantum signature other than relying on faith? Specifically, given a bit-stream, can a user verify a quantum signature in it? Researchers claim this is indecipherable and have stated a no-go theorem for post-processed bit-streams [Physical Review A \textbf{109}, 022243 (2024)]. In this article, we corroborate the results of the no-go theorem while discussing its nuances using two different random number generators and four test methods. These include the NIST statistical test suite and machine learning algorithms that strengthen the theorem. This work is relevant for companies and governments using QRNG, provided to enhance security against quantum threats.
\end{abstract}


\maketitle

\section{Introduction} \label{sec:intro}
Randomness is a well-studied topic for its applicability in many different areas ranging from randomized algorithms to cryptography. Since the seminal work of Claude Shannon \cite{shannon1948mathematical}, where he proved the information-theoretic security of one-time pad encryption by using the perfect source of randomness, its need in Quantum Key Distribution (QKD) protocols \cite{lo2005decoy} makes it an indispensable resource. The premise to avoid random bits generated by algorithms, famously known as pseudo-random number generators (PRNGs), is that they appear random due to their computational complexity. Their quality is often checked by statistical tests such as NIST Statistical Test Suite (NIST-STS) \cite{rukhin2001statistical} and Dieharder \cite{brown2018dieharder}. Despite good statistical properties, these PRNGs are vulnerable due to their lack of unpredictability. The unpredictability is typically quantified with NIST SP 800-90B and ENT. However, this unpredictable behavior is prone to advanced algorithmic attacks \cite{kelsey1998cryptanalytic}. 

The unpredictability advantage of quantum random number generators (QRNGs) is rooted in the principles of quantum mechanics, ensuring the next generated bit cannot be predicted. As per Renner's definition \cite{konig2009operational}, the quantum advantage is shown by appending a quantum metric to the typical definition of closeness to a uniform distribution. ``Proving the source of generation has a quantum origin implies that QRNGs are better than PRNGs'' is a statement that needs some experimental verification and clarification. Although a clear advantage can be seen for quantum correlations, where the correlations being quantum add privacy to these bits. However, admittedly, QRNG raw bit-streams are not statistically random, and are often post-processed using hash functions to improve their uniformity features \cite{hayashi2016more}. This is dissimilar to PRNGs, where the unpredictability is calibrated from the output bit-stream, and so is statistical independence. Once the process of generation is complete, QRNGs have a defined unpredictability metric, and the bit-stream is further enhanced for statistical independence properties of the output bit-stream by post-processing methods. The story is a little convoluted for PRNGs. Both unpredictability and uniformity in the bit-stream are measured based on the output bit-stream. Hence, post-processing methods improve statistical independence, and the unpredictability is automatically enhanced. However, when the bit-stream is ready for use in real-world applications, it is essential to have a technique to distinguish the source of origin (Quantum or Pseudo) based on the given bit-stream. The question is of paramount importance in providing access to a randomness beacon as a service. Some authors have shown a direct anti-correlation between quantumness (defined by process unpredictability) and randomness (defined by output bit statistical independence). In this article, we attempt to see if we can distinguish them on machine learning grounds. There is an advantage of QRNGs over PRNGs in terms of pattern recognition or bit-stream prediction \cite{jain2000statistical}. In the following article, we decipher this advantage of QRNGs and check whether the advantage is lost after post-processing, as verified by computational measures of randomness as suggested by the no-go theorem.

Machine Learning (ML) is a field of studying statistical models to be used by computer systems to perform tasks without explicit instructions. These models are successfully employed in diverse fields like pattern recognition \cite{jain2000statistical} and recommendation algorithms \cite{narayanan2023understanding} in everyday life. Recurrent Neural Networks (RNNs) are such a class of neural networks designed to recognize patterns in sequential data, making them suitable for tasks involving time series and prediction. Previously, RNNs have been used to predict the bit-stream of PRNG algorithms \cite{yang2018neural}. Here, we use a well-studied time series model, namely Long-Short Term Memory (LSTM) model, to predict the next bit from the previously known bit-streams. We compare its performance against NIST-STS. This paper is structured as follows. Section \ref{sec:tb} covers the techniques used to generate random numbers from algorithmic (pseudo) and quantum processes. It also discusses the post-processing method used to hash outputs. Section \ref{sec:rnd} discusses the results of PRNGs and QRNGs in both Pre and post-processed settings against the NIST-STS and the LSTM used, and finally we conclude in section \ref{sec:conc}.

\section{Theoretical Background} \label{sec:tb}
\begin{figure}[h]
    \centering
    \includegraphics[width=0.3\textwidth]{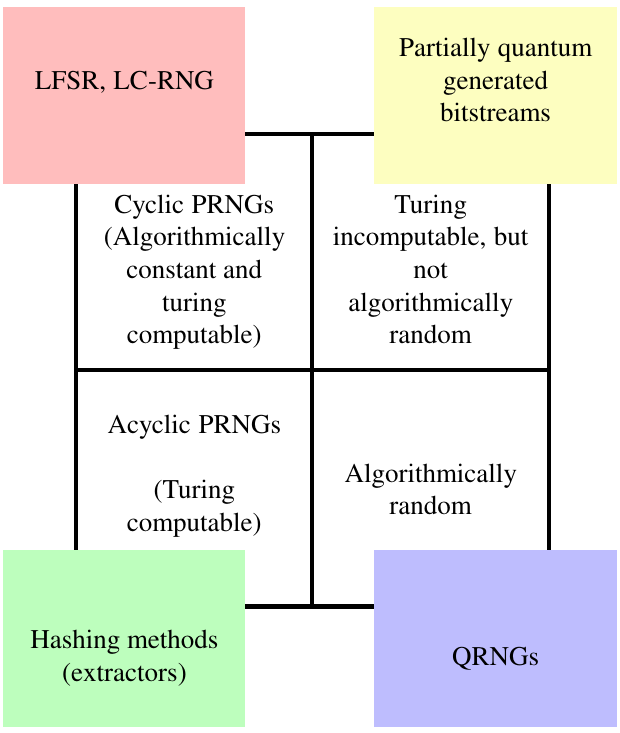}
    \caption{Classification of random number generators based on algorithmic complexity as defined by Calude \cite{calude2010experimental}.}
    \label{fig:Classification_Calude}
\end{figure}

As discussed by Calude \cite{calude2010experimental}, the classification of random numbers can be described on whether the method used is computable by a Turing machine. Such a metric could be used to differentiate PRNGs and QRNGs as the quantum correlations are not computable on a Turing machine. Amongst PRNGs, Calude differentiated between cyclic (w.r.t. periodicity) and acyclic processes, amongst other parameters of classification. Figure \ref{fig:Classification_Calude} gives a pictorial representation of the classification. Since NIST statistical hypothesis testing is aimed at deciphering bad random streams from good ones, we take bit streams from different sources, as highlighted in Fig. \ref{fig:Classification_Calude} and compare them. We go a step further by using other measures in the literature to decipher the type of source used in the generation process(quantum or pseudo) and give a supporting argument to an independently developed study providing a no-go theorem for QRNGs.

\subsection{Types of random numbers used}
Random numbers have two different faces -- statistical randomness and unpredictability of the next bit-stream based on the statistical correlations of the previous $n$ bit-streams. Any bit-stream of finite length will always have some form of statistical correlations. The goal to differentiate the two random number bit-streams with different sources of origin on grounds of statistical measures of randomness based on output bit-stream is congruous with other works \cite{li2020deep}. This leads us in the direction to differentiate bit-streams focused on the process of generation. We used Linear Congruential Generators (LCG) as our PRNG source, ChaCha20 for cryptographic RNG, and quantum entanglement-based RNG.

Linear congruential generators (LCG) are one of the fundamental building blocks in many elliptic curve cryptographic techniques \cite{gutierrez2022attacking}, and can also be used to model stochastic processes via a Markov chain model, their randomness must be studied against machine learning models to ascertain the quality of randomness generated. Under Calude's classification, stand-alone it falls under the category of cyclic PRNGs. However, post-processing it with appropriate extractors can make it acyclic. The recurrence relation generating random numbers for LCG is defined as 
\begin{equation}
    X_{n+1}=(a X_{n} + c) {\rm mod~} m,
\end{equation}
where $X_{n}$ represents the sequence of random numbers, and $m$, $a$, and $c$ are integer constants that represent the modulus, multiplier, and increment of the generator, respectively. We followed the approach of \cite{li2020deep} to vary the parameters. Stand-alone, the LCG is not a cryptographically secure PRNG. Hence, we use it to validate the model employed.

\begin{figure}[h]
    \centering
    \includegraphics[width=0.35\textwidth]{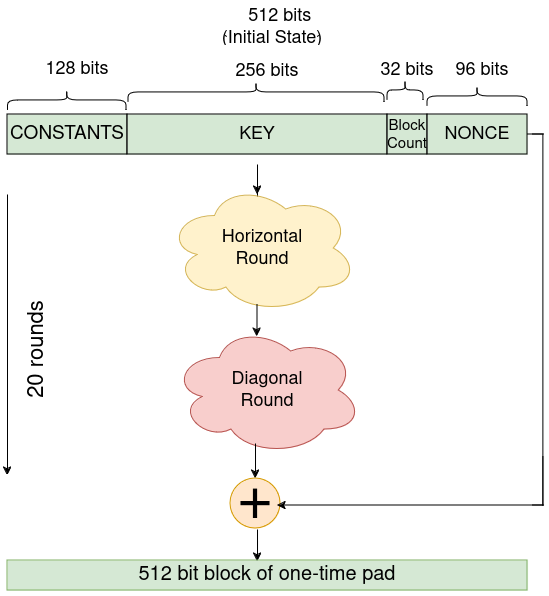}
    \caption{Working of ChaCha20 algorithm.}
    \label{fig:ChaCha20Working}
\end{figure}

For cryptographically secured PRNG (CS-PRNG), we choose ChaCha20-Poly1305, instead of its variants ChaCha8 or ChaCha12, pertaining to its active use in e-mail services. ChaCha20 performs 20 rounds of computation to achieve good unpredictability features. Each round is subdivided into four quarter rounds of tight microprocessor commands, namely an ARX cell. Here, ARX corresponds to Addition, Rotation, and XOR operations to the elements of a 4$\times$4 matrix. Each element of the matrix is 32 bits. Hence, the total matrix is written in terms of ``16$\times$32=512" bits. Figure \ref{fig:ChaCha20Working} highlights the rounds performed horizontally and diagonally. The number of rounds defines the different levels of cryptographical security. This choice is typically made to balance hardware performance and security.

For the quantum case, we use a quantum entanglement-based RNG, validated in Reference \cite{mongia2024investigating}. However, given the output bit-stream is written in a text file rather than being used directly in an application (like communication), the Bell-CHSH parameter can be used to justify the quantum origin of the source. The quantifying parameter $S$, also known as the Bell Violation parameter, is varied to two different values as demonstrated in Figure 2 of the QRNG developed \cite{mongia2024investigating}. The ChaCha20 CS-PRNG is chosen to compare to QRNG against different test suites to validate the no-go theorem. The random numbers generated from the optical hardware are a costly, delicate, and resource-intensive process. Hence, to prove their advantage, one must check for quantum correlations against standard NIST-STS and machine learning algorithms. As mentioned earlier, inevitably, the measurement of the quantum process is noisy and thus leads to classical correlations, which are post-processed.

\subsection{Randomness Extractor}
One of the best-known ways to remove unpredictable biases from an experimental setup when one cannot find the source of those biases is to redistribute them. This redistribution is often performed by operations that are one-way functions. In our case, we use a Toeplitz hash function for post-processing the random bits. It belongs to a class of functions that are two-universal \cite{tsurumaru2013dual} and makes the bit streams acyclic.

Typically, the  QRNGs, being an unpredictable source of randomness, are post-processed using the Toeplitz hash function to correct for dependent biases of the system used. To make a fair comparison, the CS-PRNG is also post-processed with same method despite passing the statistical test suite to enhance it's entropy features. This is valid as the CS-PRNG also sells itself as an unpredictable source. 

\subsection{NIST Statistical Test Suite}
NIST statistical test suite (NIST-STS) is used to check for statistical correlations in a dataset \cite{rukhin2001statistical}. Here, we employ the test suite to decipher the difference between the method of generation of the bit-stream. It contains 15 broad tests such as autocorrelation, compression, frequency, and template matching. In other words, the random numbers are tested against the chi-square hypothesis claiming the said bit-stream is random. The test static $p$-value has a threshold of 0.01.

\subsection{Machine Learning Model}
\begin{figure}[h]
    \centering
    \includegraphics[width=0.45\textwidth]{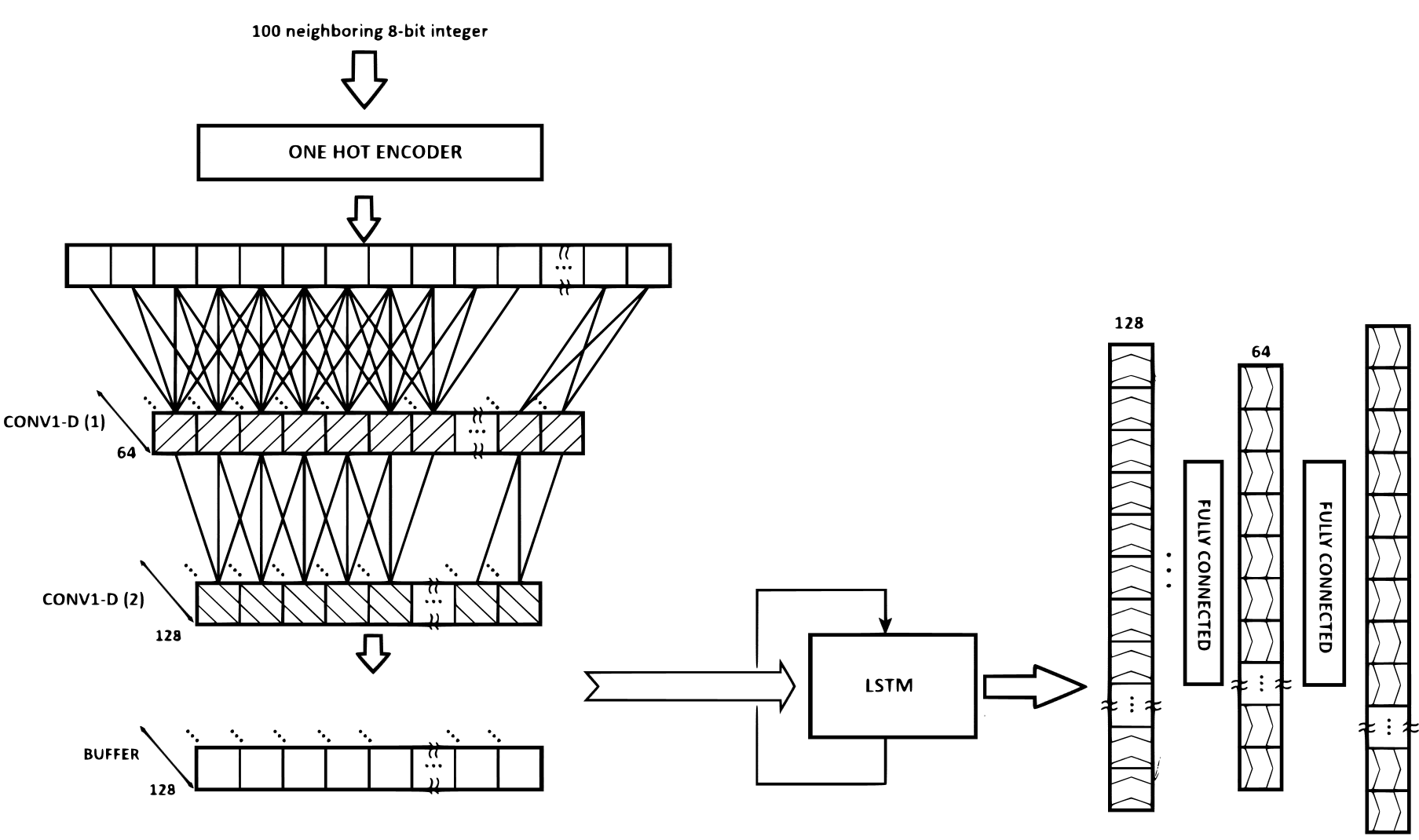}
    \caption{Working of the machine learning model.}
    \label{fig:Model_Working}
\end{figure}

The use of machine learning models to simulate time-dependent difference equations has gained significant interest due to their versatility in addressing various issues. To capture dependencies amongst the bits, we assume there are $N$ features that correspond to a single bit-stream generation. Hence, we use convolutional methods to extract these $N$ features that could correspond to the bit-stream generation. The dependencies between these features are further calculated by the LSTM model as described in the Fig. \ref{fig:Model_Working}. Finally, predictions are made based on learning the features that provide accuracy to our model. Comparing this prediction probability with guessing probability captures the advantage of our machine learning model. 

\subsubsection{Extraction of features via Convolutional Neural Networks}
Convolutional Neural Networks (CNNs) \cite{zhou2020theory} are a specialized type of neural network designed to process structured grid data, such as images. They are composed of several key components:
\begin{itemize}
    \item \textbf{Convolutional Layers}: These layers perform convolution operations on the input data, extracting features by applying filters (kernels) that slide over the input.
    \item \textbf{Activation Functions}: Following convolution, activation functions (such as ReLU or sigmoid) introduce non-linearity into the model, enabling it to learn complex patterns.
    \item \textbf{Pooling Layers}: These layers down-sample the feature maps, reducing their dimensionality while preserving essential information and discarding redundant details.
\end{itemize}

Expanding on the description of Fig. \ref{fig:Model_Working}, we input the sequence of generated random bit-streams and convert it into a string of 8-bit numbers; for this case, the baseline guess (random guess) to predict the correct next bit will be $\frac{1}{2^8} = 0.003906$. The training of the model begins with encoding $N$ (100) 8-bit integers into one-hot vectors, where each vector has all zero elements except for a single one element indicating a specific integer. These one-hot vectors, totaling 100, then undergo two convolutional layers, each followed by max-pooling of size 2. The first convolutional layer comprises 64 filters, each with a length of 5, while the second layer consists of 128 filters with a length of 3. Both convolutional layers utilize rectified linear unit (ReLU) activation functions.

After the second convolutional layer's outputs are prepared, they are fed sequentially into an LSTM layer set to produce an output size of 128. The final block of the LSTM output contains comprehensive sequence information. This output of size 128 connects to two fully connected layers, each employing sigmoid and softmax functions as activation functions. These fully connected layers have output sizes of 64 and 256, respectively.

\subsubsection{LSTM Model}
\begin{figure}[h]
    \centering
    \includegraphics[width=0.45\textwidth]{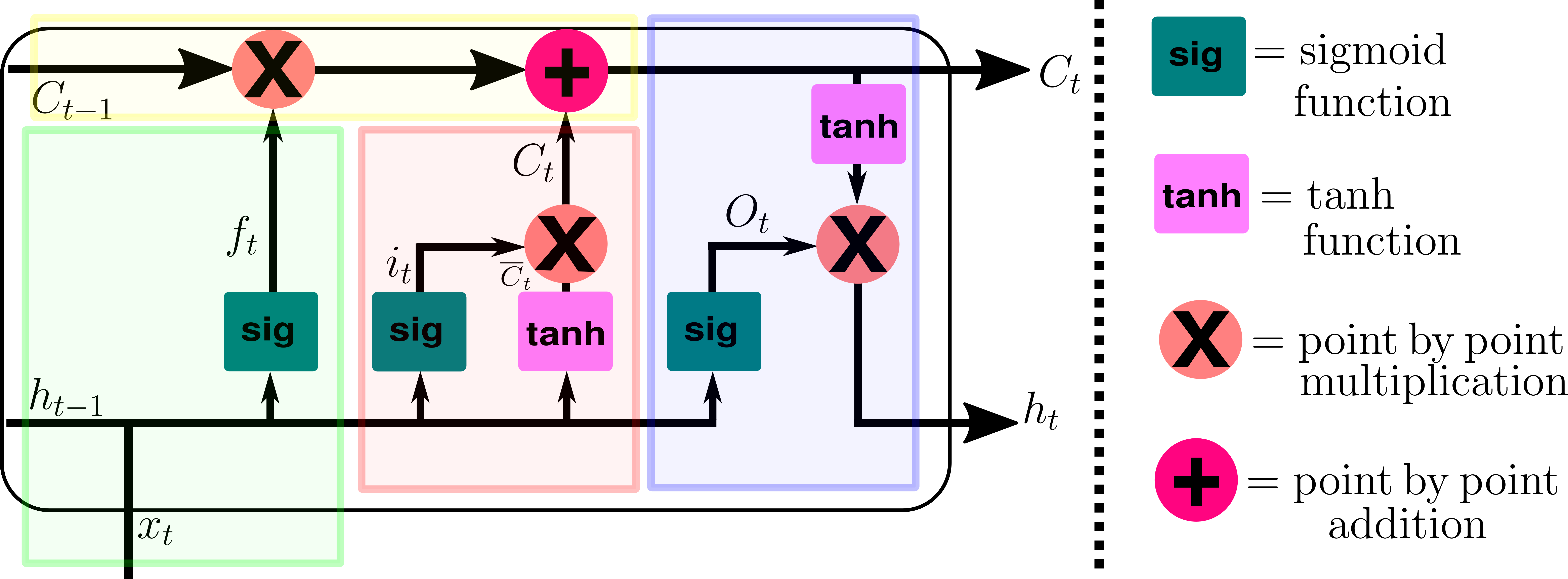}
    \caption{Detailed working of an LSTM cell: Green box represents the forget gate, red box represents the input gate, blue ox represents output gate, yellow box represents the cell state: operations performed on the input state are labeled on the right side of the dashed line.}
    \label{fig:LSTM_Working}
\end{figure}

As the name suggests, LSTM looks out for both long-term and short-term temporal dependencies using laws of differentiation. This is a better choice of temporal modeling compared to conventional recurrent neural networks (RNNs) solving the vanishing gradient problem with an extra cell for long-term context. LSTM is the basic building block of its model. The cells concatenated horizontally form the information highway for context where the output of one cell is input to the next cell. In this section, we describe its working in detail, which is also illustrated in Fig. \ref{fig:LSTM_Working}. Since we are interested in catching temporal dependencies, we need to understand how the context (previous bits in our case) based prediction happens. For contextual information to be available at hand, we look into how this context is stored in the long and short-term memories of the LSTM cell. Long-term stores context for longer temporal correlations between the bit-stream, while the short-term stores context for the last few recurring bits. However, both long- and short-term cells influence each other. All this working of the time-series forecasting model can be understood in terms of the simplest recurring unit of LSTM, the LSTM cell. Succinctly, it is an information highway for context (biases of the optical equipment used reflected in previous bits) to find long-term and short-term dependencies of the $n^{th}$ bit-stream on previous bit-streams. Now we focus on how the long-term context is stored in the memory of the LSTM cell.

The cell state $C_{t-1}$, in Fig. \ref{fig:LSTM_Working}, represents the information stored in long-term memory. The hidden state $h_{t-1}$ depicts the information stored in the short-term memory. For the next cell with new input information as $x_{t}$, a non-linear activation of the sigmoid function is applied to the cell state based on the input hidden state $h_{t-1}$. This activation is zero if $x_{t}$ provides no new information compared to $h_{t-1}$ and one otherwise. Thus, this activation updates the cell state, $C_{t-1}$ providing the amount of relevance of the new information $x_{t}$ and $h_{t-1}$. Once the long-term memory, the cell state $C_{t-1}$ is updated, the cell state is transferred onto the next step. Here, the input to the cell state is defined pertaining to the relevance of the new input information, $x_{t}$. After the input has been updated, the next part of the LSTM cell, the output gate, does two things. One, it generates the output for the next bit prediction, $y_{t}$. Secondly, it updates the short-term memory $h_{t}$ based on the $h_{t-1}$, $x_{t}$, and $C_{t}$ through appropriate activations which serve as input to the next recurring gate.

\subsection{Algorithmic measures of randomness}
Algorithmic randomness of a bit-stream focuses on the computability of the algorithm on a universal Turing machine. The differentiating factor amongst the PRNGs and QRNGs reduces the problem of computability. For example, the well-known RSA algorithm used in cryptography trades hardness for randomness between computational complexity classes BPP and P. One such measure of randomness is Kolmogorov Complexity (a measure of incomputability). We use a lossless LZ-76 compression algorithm to calculate the Kolmogorov Complexity \cite{lempel2003complexity}. Heuristically, the more the compression, the lesser the randomness in a bit-stream. 

\subsubsection{Kolmogorov Complexity}
Kolmogorov complexity is a measure of computability. Kolmogorov Complexity is defined as the length of the shortest program to mimic the computation of a given program. Although it is a theoretical concept based on deciding the halting problem, we follow a pragmatic approach to approximate it using the compression technique. It is a theoretical concept in computer science based on Turing computability. The Kolmogorov complexity $K(x)$ of a string $x$ is defined as 
\begin{equation}
    K(x) = \min \left\{ \left| p \right| \mid U(p) = x \right\}
\end{equation}
where $U$ is a universal Turing machine, $p$ is a program, and $\left| p \right|$ is the length of the program. A string $x$ is incompressible or Kolmogorov random if $K(x) \geq \left| x \right|$. As an example, consider the bit-stream, ``0101010101010101''. The typical and shortest method to describe the text, excluding common overheads, can be easily verified as ``01'' $\times 8$. Practically, the Kolmogorov complexity can be approximated by using compression algorithms. This approximation also limits our case of deciphering QRNGs from PRNGs and reduces the efficacy of all algorithms to efficiently calculable measures as stated in the no-go theorem \cite{tsurumaru2024indistinguishability}. Amongst the LZ-compression family, we specifically focus on the predecessor of all these algorithms, the LZ-76 algorithm. A string with high Kolmogorov complexity is considered random or incompressible, while a string with low complexity can be described succinctly.

The length of the output produced by LZ-76 can be seen as an upper bound for the Kolmogorov complexity of the input string \cite{li2020deep}. Specifically, if $x$ is compressed using LZ-76 to produce a string $y$, then:
\begin{equation}
    K(x) \leq \left| y \right| + O(\log y).
\end{equation}
The algorithm dynamically builds a dictionary as it processes the input bit-stream, allowing it to adapt to varying bit-stream patterns as discussed in  \cite{williams1991extremely}.

\subsubsection{Borel Normality}
Borel Normality is a measure that checks for Independent and Identically Distributed (IID) criteria in the case of random numbers. Mathematically, it can be expressed as
\begin{equation}
    \left| \frac{N_j^m(s^N)}{|s^N|_m}-\frac{1}{2^m} \right| \leq \frac{1}{\log_2 |s^N|},
\end{equation}
where $s$ is the bit-string of length $N$ and $m$ is the level of hierarchy at which we check for IID criteria and $N_j^m(s^N)$ is the number of events for a particular combination $j$ in a $m$-bit hierarchy. Additionally, if a distribution from a random number generation process (quantum or pseudo) follows IID criteria, the assumptions on the randomness extractor can be relaxed \cite{TCS-010}.

\section{Results and Discussion} \label{sec:rnd}
The results are organized in Table \ref{tab:theoretical_measures}. 
\begin{table}[h]
    \centering
    \begin{tabular}{p{3.25cm}|l|l}
        \textbf{Measures} & \textbf{Pre-processed} & \textbf{Post-processed} \\ \hline \hline
        \textbf{Statistical randomness: NIST-STS} & Case I & Case II \\ \hline
        \textbf{Unpredictability: LSTM} & Case III & Case IV \\ \hline
        \textbf{Algorithmic randomness} & Pre-processed & Post-processed \\ \hline
        \textbf{Kolmogorov complexity} & Case V & Case V \\ \hline
        \textbf{Borel normality} & Case VI & Case VI \\ \hline
    \end{tabular}
    \caption{Comparison of randomness measures in pre-processed and post-processed scenarios.}
    \label{tab:theoretical_measures}
\end{table}

\subsection{Case I: Pre-processed PRNG and QRNG bit-streams against NIST-STS}
\begin{figure}[h]
    \centering
    \includegraphics[width=0.45\textwidth]{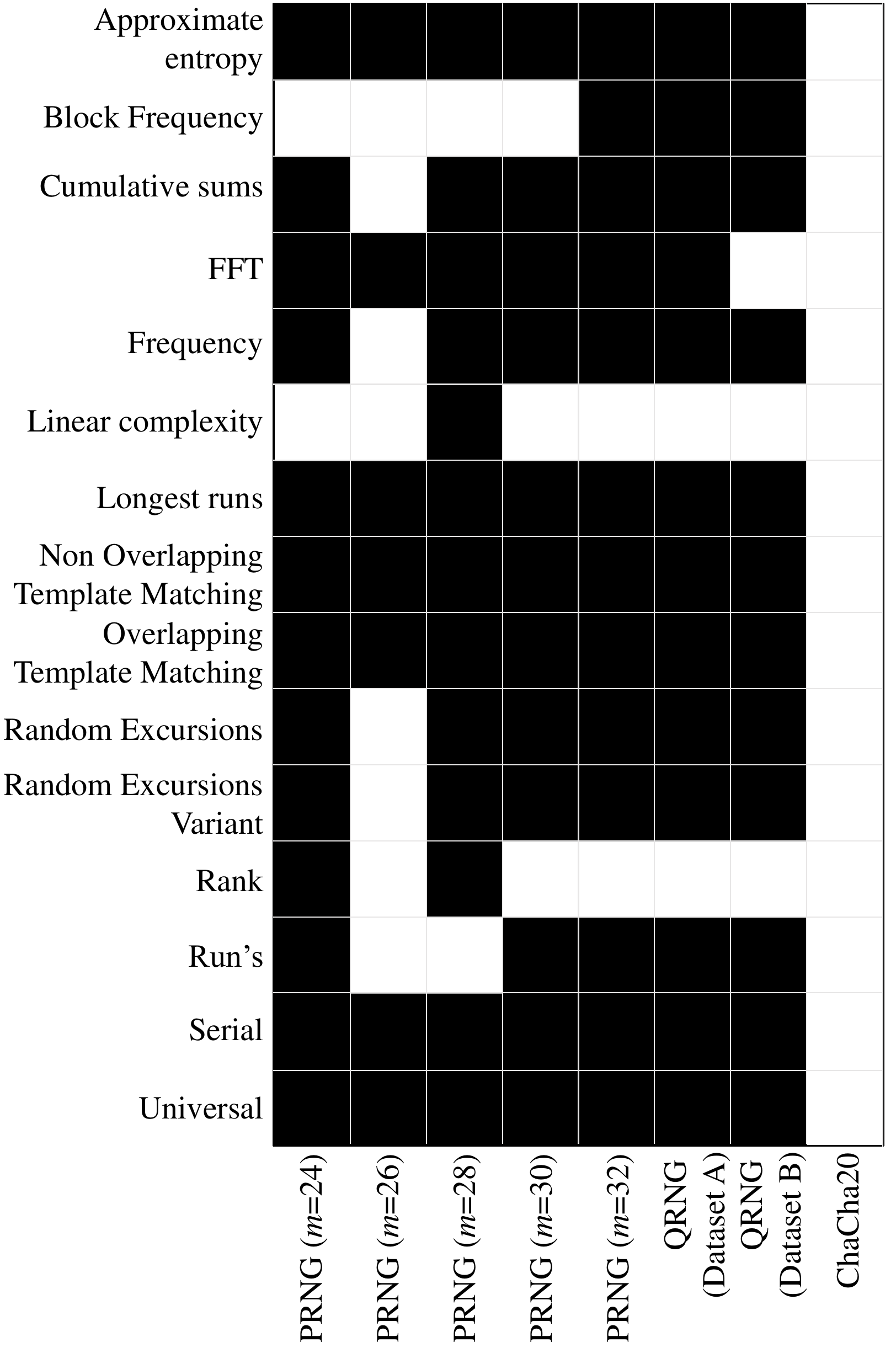}
    \caption{NIST-STS results for pre-processed bit-stream of length 5 M; only ChaCha20 passes all 15 tests. Black (white) color represent the fail (pass) in the individual tests.}
    \label{fig:Pre-processed_NIST}
\end{figure}

To compare PRNGs and QRNGs, we choose an initial dataset of 5 million bit-streams. Initial testing on the raw dataset for randomness of PRNGs shows a failure in multiple sub-tests, as shown in Fig. \ref{fig:Pre-processed_NIST}. One can notice that the block frequency test shows a completely random sequence while the frequency test fails. This suggests that the raw bits from both PRNGs are weak sources of randomness. For QRNGs, this is an advantage that at least no two relative tests contradict each other. Also, no clear indication is seen in the test static $p$-value which could act as a differentiator amongst them.

\subsection{Case II: Post-processed PRNG and QRNG (of length 1.2M) bit-streams against NIST-STS}
\begin{figure}[h]
    \centering
    \includegraphics[width=0.45\textwidth]{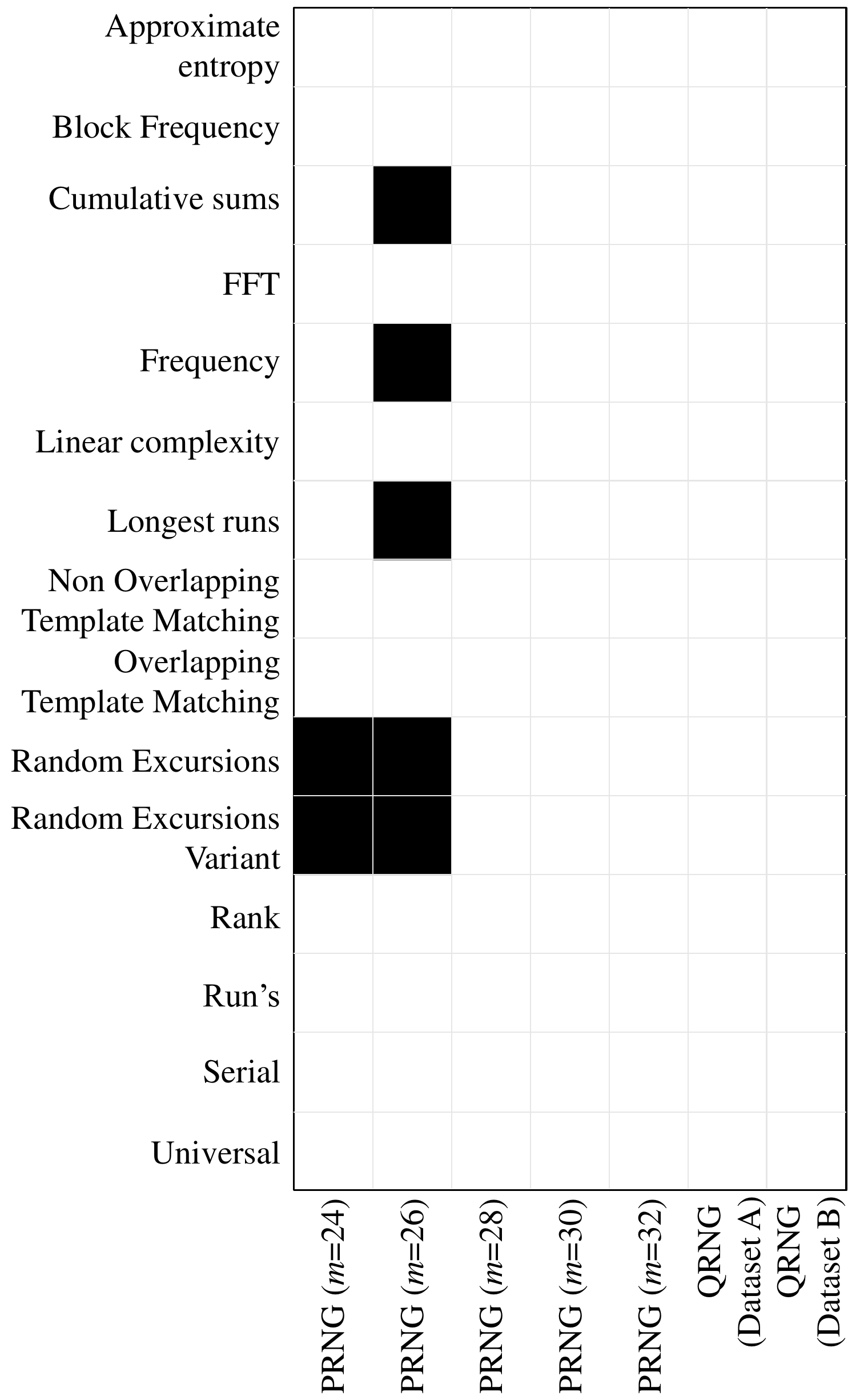}
    \caption{NIST-STS results for post-processed bit-stream of length 1.2 M. Black or white color represent the fail or pass in individual tests, respectively.}
    \label{fig:Post-Processed_NIST}
\end{figure}

To extract randomness, we post-process the raw bit-streams (5 M) to a hashed length of 1.2 M using the Toeplitz matrix multiplication. The hashed output bits are tested against the NIST-STS for patterns. Exact matching of frequency and linear complexity tests for the QRNG dataset is an indicator that the data has been heavily post-processed. However, 2-universality of the Toeplitz hash function ensures that such post-processed PRNGs are statistically intractable. The results are shown in Fig. \ref{fig:Post-Processed_NIST}.

\subsection{Case III: Pre-processed PRNG and QRNG against LSTM model}
PRNGs and QRNGs on the pre-processed bit-stream don't have good statistical properties. However, their unpredictability is a parameter that differentiates them but is not addressed in this article. The results are highlighted in Table \ref{tab:ML_QRNG_ChaCha20}. Here, we see that there is a high prediction probability among PRNGs under almost all variations of $m$, $a$, and $c$. This is a noticeable difference between PRNGs and QRNGs.
\begin{table}[h]
    \centering
    \begin{tabular}{p{2.5cm}|l|l}
        \textbf{Tests} & $P_{ml} a_1c_1$ & $P_{ml} a_2c_2$ \\\hline \hline
        \textbf{PRNG} ($m$=24) & 0.327\% & 71.331\% \\ \hline
        \textbf{PRNG} ($m$=26) & 89.076\% & 87.184\% \\ \hline
        \textbf{PRNG} ($m$=28) & 95.563\% & 73.537\% \\ \hline
        \textbf{PRNG} ($m$=30) & 65.938\% & 51.408\% \\ \hline
        \textbf{PRNG} ($m$=32) & 59.958\% & 1.605\% \\ \hline
    \end{tabular}
    \caption{Machine Learning Model trained on a known PRNG (LCG) for model verification.}
    \label{tab:ML_LCG_Pre}
\end{table}

\begin{table}[h]
    \centering
    \begin{tabular}{l|p{2.5cm}|p{2.5cm}|p{2cm}}
        \textbf{Tests} & \textbf{QRNG (Dataset A)} & \textbf{QRNG (Dataset B)} & \textbf{ChaCha20} \\ \hline \hline
        $P_{ml}$ & 0.463\% & 0.414\% & 6.038\% \\ \hline
        $P_{g}$ & 0.3906\% & 0.3906\% & 0.3906\% \\ \hline
    \end{tabular}
    \caption{Testing Pre-processed Quantum RNG and ChaCha20 (CS-PRNG) against machine learning model, $P_{ml}:$ next bit prediction probability by machine learning model and $P_{g}:$ next bit prediction probability on random guessing ($=\frac{1}{2^8}$).}
    \label{tab:ML_QRNG_ChaCha20}
\end{table}

\subsection{Case IV: Post-processed PRNG and QRNG against LSTM model}
The post-processed bit-stream is checked for patterns against the LSTM model whose results are highlighted in \ref{tab:LSTM_PostProcessed_Training_Data}. One can see that the probability of detecting patterns via the model is close to the guessing probability (0.391\%) in all the cases irrespective of the variation of $m$ or the length of variation of the post-processed bit-stream as shown in Table \ref{tab:LSTM_PostProcessed_Training_Data}. From the data, one can infer that once post-processed, the hashing methods are stronger than the computational measurement techniques used to predict PRNGs.  
\begin{table}[h]
    \centering
    \begin{tabular}{p{2.4cm}|l|l|l|l}
        \textbf{Tests} & $P_{ml}l_{1 M}$ & $P_{ml}l_{1.2 M}$ & $P_{ml}l_{1.5 M}$ & $P_{ml}l_{2.0 M}$ \\ \hline \hline
        \textbf{PRNG} ($m$=24) & 0.494\% & 0.351\% & 0.361\% & 0.403\% \\ \hline
        \textbf{PRNG} ($m$=26) & 0.458\% & 0.472\% & 0.385\% & 0.331\% \\ \hline
        \textbf{PRNG} ($m$=28) & 0.410\% & 0.361\% & 0.313\% & 0.439\% \\ \hline
        \textbf{PRNG} ($m$=30) & 0.361\% & 0.301\% & 0.409\% & 0.403\% \\ \hline
        \textbf{PRNG} ($m$=32) & 0.373\% & 0.371\%& 0.385\%& 0.409\% \\ \hline
    \end{tabular}
    \caption{Variation of post-processing length for LC-RNG against machine learning methods.}
    \label{tab:LSTM_PostProcessed_Training_Data}
\end{table}

Also, one is unable to differentiate between PRNGs and QRNGs as shown in Table \ref{tab:LSTM_PostProcessed_Actual_Data}.  This is a clear indicator that the post-processing method used (hashing) masks the quantum nature against computational measures. In terms of complexity, if we measure the properties of QRNGs and PRNGs against polynomial time measures, both classes appear to be similar. The results provide evidence that class BQP is larger than class BPP as shown by unprocessed QRNGs being unpredictable against machine learning models compared to unprocessed PRNGs. For the post-processed data, both classes BQP and BPP appear equivalent from a computational stand-point.

\begin{table}[h]
    \centering
    \begin{tabular}{p{2.4cm}|l|l|l|l}
        \textbf{Tests} & $P_{ml}l_{1 M}$ & $P_{ml}l_{1.2 M}$ & $P_{ml}l_{1.5 M}$ & $P_{ml}l_{2.0 M}$ \\ \hline \hline
        \textbf{QRNG (Dataset A)} & 0.470\% & 0.311\% & 0.311\% & 0.337\% \\ \hline
        \textbf{QRNG (Dataset B)} & 0.386\% & 0.391\% & 0.391\% & 0.307\% \\ \hline
        \textbf{ChaCha20} & 0.518\% & 0.351\% & 0.351\% & 0.563\% \\\hline
    \end{tabular}
    \caption{Variation of post-processing length for LC-RNG against machine learning methods.}
    \label{tab:LSTM_PostProcessed_Actual_Data}
\end{table}

\subsection{Case V: Kolmogorov Complexity for PRNG and QRNG pre and post-processed bit-stream}
The Kolmogorov Complexity of both pre- and post-processed bit-streams for PRNGs and QRNGs is highlighted in Table \ref{tab:kol_Preand-post-processed}. Here, all values of Kolmogorov complexity are shown w.r.t. the seed used in the randomness extractor. The Kolmogorov complexity of the seed (borrowed from MT-19937) is 1.382 \cite{matsumoto1998mersenne}. All values shown below are normalized w.r.t. the quality of the seed used for extraction. It can be inferred from Table \ref{tab:kol_Preand-post-processed} that once post-processed heavily, everything becomes indecipherable. 
\begin{table}[h]
    \centering
    \begin{tabular}{p{2.0cm}|p{1.5cm}|p{1.5cm}|p{1.6cm}|p{1.6cm}}
        \textbf{Files} & \textbf{Pre- processed} & \textbf{Pre- processed} & \textbf{Post- processed} & \textbf{Post- processed} \\ \hline \hline
        \textbf{PRNG} ($m$=24) & 0.803 & 0.907 & 1 & 1 \\ \hline
        \textbf{PRNG} ($m$=26) & 0.868 & 0.799 & 1 & 1 \\ \hline
        \textbf{PRNG} ($m$=28) & 0.519 & 0.910 & 1 & 1 \\ \hline
        \textbf{PRNG} ($m$=30) & 0.896 & 0.936 & 1  & 1 \\ \hline
        \textbf{PRNG} ($m$=32) & 0.910 & 0.493 & 1 & 1 \\ \hline
        \textbf{QRNG (Dataset A)} & 0.975 & N.A. & 1 & N.A. \\ \hline
        \textbf{QRNG (Dataset B)} & 0.976 & N.A. & 1 & N.A. \\ \hline
        \textbf{ChaCha20} & 0.976 & N.A. & 1 & N.A. \\ \hline
    \end{tabular}
    \caption{Kolmogorov Complexity calculated via LZ compression.}
    \label{tab:kol_Preand-post-processed}
\end{table}

\subsection{Case VI: Borel Normality for PRNG and QRNG pre and post-processed bit-stream}
The Borel Normality of both -processed bit-streams for PRNGs and QRNGs is highlighted in Table \ref{tab:Unprocessed_Borelnormality}. While the post-processed bit-streams for PRNGs and QRNGs successfully pass the test for all cases, we notice the drastic change in Borel normality criteria for QRNGs for pre-processed and post-processed bit-streams. This is intuitive as well; the more unpredictable the bit-stream, the less it should adhere to uniform distribution and thus, IID criteria.

\begin{table}[h]
    \centering
    \begin{tabular}{l|l|l|l|l}
        \textbf{Number of bit-streams} & \textbf{1-bit} & \textbf{2-bit} & \textbf{3-bit} & \textbf{4-bit} \\ \hline \hline
        \textbf{PRNG} ($m$=24) & 1 & 0 & 0 & 0 \\ \hline
        \textbf{PRNG} ($m$=26) & 1 & 1 & 1 & 0 \\ \hline
        \textbf{PRNG} ($m$=28) & 1 & 1 & 1 & 1 \\ \hline
        \textbf{PRNG} ($m$=30) & 0 & 1 & 1 & 1 \\ \hline
        \textbf{QRNG (Dataset A)} & 0 & 0 & 0 & 0 \\ \hline
        \textbf{QRNG (Dataset B)} & 0 & 0 & 0 & 0 \\ \hline
        \textbf{ChaCha20} & 1 & 1 & 1 & 1 \\ \hline
    \end{tabular}
    \caption{Borel Normality Tests on pre-processed Data.}
    \label{tab:Unprocessed_Borelnormality}
\end{table}

For the post-processed bit-streams, one can clearly see that all the bit-streams, irrespective of their origin (PRNG, QRNG, or CS-PRNG) pass the Borel normality criteria as shown in table \ref{tab:Post-processed_Borelnormality}. This is the fourth measure that supports the claims made by the no-go theorem. The drastic change in QRNGs following IID criteria after being post-processed provides empirical evidence that post-processing helps us improve the uniformity features of quantum random number generators. 

\begin{table}[h]
    \centering
    \begin{tabular}{l|l|l|l|l}
        \textbf{Number of bit-streams} & \textbf{1-bit} & \textbf{2-bit} & \textbf{3-bit} & \textbf{4-bit} \\ \hline \hline
        \textbf{PRNG} ($m$=24) & 1 & 1 & 1 & 1 \\ \hline
        \textbf{PRNG} ($m$=26) & 1 & 1 & 1 & 1 \\ \hline
        \textbf{PRNG} ($m$=28) & 1 & 1 & 1 & 1 \\ \hline
        \textbf{PRNG} ($m$=30) & 1 & 1 & 1 & 1 \\ \hline
        \textbf{QRNG (Dataset A)} & 1 & 1 & 1 & 1 \\ \hline
        \textbf{QRNG (Dataset B)} & 1 & 1 & 1 & 1 \\ \hline
        \textbf{ChaCha20} & 1 & 1 & 1 & 1 \\ \hline
    \end{tabular}
    \caption{Borel Normality Tests on Post-processed Data.}
    \label{tab:Post-processed_Borelnormality}
\end{table}

\section{Conclusion} \label{sec:conc}
We provide three major results with this study. Firstly, we provide four independent evidences of a no-go theorem against a chosen QRNG and PRNG. Typically, the initial work stating the no-go theorem \cite{tsurumaru2024indistinguishability} considered only two measures, namely, Kolmogorov complexity and Borel normality conditions. We reproduce similar conclusions with an additional statement referring to where exactly the definition of effectively calculable measures constraint comes into the picture. We have appended their conclusions in a more rigorous manner from a practical standpoint using both NIST-STS and ML models. In an ideal case, one should use multiple hybridizations of space-time complexity models where the space model extracts the features out of the bit-stream while the time model finds out dependencies between the features so that our negative results are model-agnostic. Secondly, we see that Chacha20 shows weakness against machine learning models despite the unprocessed bit-stream passing NIST-STS. This weakness could be further explored with advanced machine learning algorithms. Thirdly, we see how quantum unpredictability anti-correlates to computational standards (such as IID).

To claim an advantage of QRNGs over PRNGs, QRNGs are unpredictable as they couldn't be caught by machine learning algorithms. However, they cannot be used without post-processing as they fail NIST-STS. After post-processing, one is unable to decipher between the QRNG and PRNG from the bit-stream. In the fundamental directions and from a quantum complexity standpoint, the question of deciphering QRNGs from PRNGs boils down to their usage in complexity classes BQP and BPP. From the work of Vazirani \cite{Vazirani_1993_QCT}, it is known that BPP lies in BQP. We point out that since the bit-streams are post-processed, there appears to be a reduction from quantum complexity class to computational complexity class for QRNGs (which lie in BQP because of quantum entanglement methods) and they become comparable to CS-PRNGs. To make a stronger claim on the reduction, more verification is required, say with transformers on the machine learning front, as the context window for transformers is larger than LSTM models. 

\section*{Acknowledgments}
SP acknowledges the support of the Department of Space, Government of India. VM acknowledges fruitful discussions with Prof. Ravi Hegde and Dr. Soumyashree Panda.

\section*{Data Availability}
The data that support the findings of this study are available from the corresponding author upon reasonable request.

\section*{Disclosures}
The authors declare that they have no conflicts of interest related to this article.

\bibliography{manuscript}

\begin{thebibliography}{22}%
\makeatletter
\providecommand \@ifxundefined [1]{%
 \@ifx{#1\undefined}
}%
\providecommand \@ifnum [1]{%
 \ifnum #1\expandafter \@firstoftwo
 \else \expandafter \@secondoftwo
 \fi
}%
\providecommand \@ifx [1]{%
 \ifx #1\expandafter \@firstoftwo
 \else \expandafter \@secondoftwo
 \fi
}%
\providecommand \natexlab [1]{#1}%
\providecommand \enquote  [1]{``#1''}%
\providecommand \bibnamefont  [1]{#1}%
\providecommand \bibfnamefont [1]{#1}%
\providecommand \citenamefont [1]{#1}%
\providecommand \href@noop [0]{\@secondoftwo}%
\providecommand \href [0]{\begingroup \@sanitize@url \@href}%
\providecommand \@href[1]{\@@startlink{#1}\@@href}%
\providecommand \@@href[1]{\endgroup#1\@@endlink}%
\providecommand \@sanitize@url [0]{\catcode `\\12\catcode `\$12\catcode `\&12\catcode `\#12\catcode `\^12\catcode `\_12\catcode `\%12\relax}%
\providecommand \@@startlink[1]{}%
\providecommand \@@endlink[0]{}%
\providecommand \url  [0]{\begingroup\@sanitize@url \@url }%
\providecommand \@url [1]{\endgroup\@href {#1}{\urlprefix }}%
\providecommand \urlprefix  [0]{URL }%
\providecommand \Eprint [0]{\href }%
\providecommand \doibase [0]{https://doi.org/}%
\providecommand \selectlanguage [0]{\@gobble}%
\providecommand \bibinfo  [0]{\@secondoftwo}%
\providecommand \bibfield  [0]{\@secondoftwo}%
\providecommand \translation [1]{[#1]}%
\providecommand \BibitemOpen [0]{}%
\providecommand \bibitemStop [0]{}%
\providecommand \bibitemNoStop [0]{.\EOS\space}%
\providecommand \EOS [0]{\spacefactor3000\relax}%
\providecommand \BibitemShut  [1]{\csname bibitem#1\endcsname}%
\let\auto@bib@innerbib\@empty
\bibitem [{\citenamefont {Shannon}(1948)}]{shannon1948mathematical}%
  \BibitemOpen
  \bibfield  {author} {\bibinfo {author} {\bibfnamefont {C.~E.}\ \bibnamefont {Shannon}},\ }\bibfield  {title} {\bibinfo {title} {A mathematical theory of communication},\ }\href {https://doi.org/10.1002/j.1538-7305.1948.tb01338.x} {\bibfield  {journal} {\bibinfo  {journal} {The Bell system technical journal}\ }\textbf {\bibinfo {volume} {27}},\ \bibinfo {pages} {379} (\bibinfo {year} {1948})}\BibitemShut {NoStop}%
\bibitem [{\citenamefont {Lo}\ \emph {et~al.}(2005)\citenamefont {Lo}, \citenamefont {Ma},\ and\ \citenamefont {Chen}}]{lo2005decoy}%
  \BibitemOpen
  \bibfield  {author} {\bibinfo {author} {\bibfnamefont {H.-K.}\ \bibnamefont {Lo}}, \bibinfo {author} {\bibfnamefont {X.}~\bibnamefont {Ma}},\ and\ \bibinfo {author} {\bibfnamefont {K.}~\bibnamefont {Chen}},\ }\bibfield  {title} {\bibinfo {title} {Decoy state quantum key distribution},\ }\href {https://doi.org/10.1103/PhysRevLett.94.230504} {\bibfield  {journal} {\bibinfo  {journal} {Physical Review Letters}\ }\textbf {\bibinfo {volume} {94}},\ \bibinfo {pages} {230504} (\bibinfo {year} {2005})}\BibitemShut {NoStop}%
\bibitem [{\citenamefont {Rukhin}\ \emph {et~al.}(2001)\citenamefont {Rukhin}, \citenamefont {Soto}, \citenamefont {Nechvatal}, \citenamefont {Smid}, \citenamefont {Barker}, \citenamefont {Leigh}, \citenamefont {Levenson}, \citenamefont {Vangel}, \citenamefont {Banks}, \citenamefont {Heckert} \emph {et~al.}}]{rukhin2001statistical}%
  \BibitemOpen
  \bibfield  {author} {\bibinfo {author} {\bibfnamefont {A.}~\bibnamefont {Rukhin}}, \bibinfo {author} {\bibfnamefont {J.}~\bibnamefont {Soto}}, \bibinfo {author} {\bibfnamefont {J.}~\bibnamefont {Nechvatal}}, \bibinfo {author} {\bibfnamefont {M.}~\bibnamefont {Smid}}, \bibinfo {author} {\bibfnamefont {E.}~\bibnamefont {Barker}}, \bibinfo {author} {\bibfnamefont {S.}~\bibnamefont {Leigh}}, \bibinfo {author} {\bibfnamefont {M.}~\bibnamefont {Levenson}}, \bibinfo {author} {\bibfnamefont {M.}~\bibnamefont {Vangel}}, \bibinfo {author} {\bibfnamefont {D.}~\bibnamefont {Banks}}, \bibinfo {author} {\bibfnamefont {A.}~\bibnamefont {Heckert}}, \emph {et~al.},\ }\href {https://doi.org/10.6028/NIST.SP.800-22r1a} {\emph {\bibinfo {title} {A statistical test suite for random and pseudorandom number generators for cryptographic applications}}},\ Vol.~\bibinfo {volume} {22}\ (\bibinfo  {publisher} {US Department of Commerce, Technology Administration, National Institute of},\ \bibinfo {year} {2001})\BibitemShut {NoStop}%
\bibitem [{\citenamefont {Brown}\ \emph {et~al.}(2018)\citenamefont {Brown}, \citenamefont {Eddelbuettel},\ and\ \citenamefont {Bauer}}]{brown2018dieharder}%
  \BibitemOpen
  \bibfield  {author} {\bibinfo {author} {\bibfnamefont {R.~G.}\ \bibnamefont {Brown}}, \bibinfo {author} {\bibfnamefont {D.}~\bibnamefont {Eddelbuettel}},\ and\ \bibinfo {author} {\bibfnamefont {D.}~\bibnamefont {Bauer}},\ }\bibfield  {title} {\bibinfo {title} {Dieharder},\ }\href {https://webhome.phy.duke.edu/~rgb/General/dieharder.php} {\bibfield  {journal} {\bibinfo  {journal} {Duke University Physics Department Durham, NC}\ ,\ \bibinfo {pages} {27708}} (\bibinfo {year} {2018})}\BibitemShut {NoStop}%
\bibitem [{\citenamefont {Kelsey}\ \emph {et~al.}(1998)\citenamefont {Kelsey}, \citenamefont {Schneier}, \citenamefont {Wagner},\ and\ \citenamefont {Hall}}]{kelsey1998cryptanalytic}%
  \BibitemOpen
  \bibfield  {author} {\bibinfo {author} {\bibfnamefont {J.}~\bibnamefont {Kelsey}}, \bibinfo {author} {\bibfnamefont {B.}~\bibnamefont {Schneier}}, \bibinfo {author} {\bibfnamefont {D.}~\bibnamefont {Wagner}},\ and\ \bibinfo {author} {\bibfnamefont {C.}~\bibnamefont {Hall}},\ }\bibfield  {title} {\bibinfo {title} {Cryptanalytic attacks on pseudorandom number generators},\ }in\ \href {https://doi.org/10.1007/3-540-69710-1_12} {\emph {\bibinfo {booktitle} {International workshop on fast software encryption}}}\ (\bibinfo {organization} {Springer},\ \bibinfo {year} {1998})\ pp.\ \bibinfo {pages} {168--188}\BibitemShut {NoStop}%
\bibitem [{\citenamefont {Konig}\ \emph {et~al.}(2009)\citenamefont {Konig}, \citenamefont {Renner},\ and\ \citenamefont {Schaffner}}]{konig2009operational}%
  \BibitemOpen
  \bibfield  {author} {\bibinfo {author} {\bibfnamefont {R.}~\bibnamefont {Konig}}, \bibinfo {author} {\bibfnamefont {R.}~\bibnamefont {Renner}},\ and\ \bibinfo {author} {\bibfnamefont {C.}~\bibnamefont {Schaffner}},\ }\bibfield  {title} {\bibinfo {title} {The operational meaning of min-and max-entropy},\ }\href {https://doi.org/10.1109/TIT.2009.2025545} {\bibfield  {journal} {\bibinfo  {journal} {IEEE Transactions on Information theory}\ }\textbf {\bibinfo {volume} {55}},\ \bibinfo {pages} {4337} (\bibinfo {year} {2009})}\BibitemShut {NoStop}%
\bibitem [{\citenamefont {Hayashi}\ and\ \citenamefont {Tsurumaru}(2016)}]{hayashi2016more}%
  \BibitemOpen
  \bibfield  {author} {\bibinfo {author} {\bibfnamefont {M.}~\bibnamefont {Hayashi}}\ and\ \bibinfo {author} {\bibfnamefont {T.}~\bibnamefont {Tsurumaru}},\ }\bibfield  {title} {\bibinfo {title} {More efficient privacy amplification with less random seeds via dual universal hash function},\ }\href {https://doi.org/10.1109/TIT.2016.2526018} {\bibfield  {journal} {\bibinfo  {journal} {IEEE Transactions on Information Theory}\ }\textbf {\bibinfo {volume} {62}},\ \bibinfo {pages} {2213} (\bibinfo {year} {2016})}\BibitemShut {NoStop}%
\bibitem [{\citenamefont {Jain}\ \emph {et~al.}(2000)\citenamefont {Jain}, \citenamefont {Duin},\ and\ \citenamefont {Mao}}]{jain2000statistical}%
  \BibitemOpen
  \bibfield  {author} {\bibinfo {author} {\bibfnamefont {A.~K.}\ \bibnamefont {Jain}}, \bibinfo {author} {\bibfnamefont {R.~P.~W.}\ \bibnamefont {Duin}},\ and\ \bibinfo {author} {\bibfnamefont {J.}~\bibnamefont {Mao}},\ }\bibfield  {title} {\bibinfo {title} {Statistical pattern recognition: A review},\ }\href {https://doi.org/10.1109/34.824819} {\bibfield  {journal} {\bibinfo  {journal} {IEEE Transactions on pattern analysis and machine intelligence}\ }\textbf {\bibinfo {volume} {22}},\ \bibinfo {pages} {4} (\bibinfo {year} {2000})}\BibitemShut {NoStop}%
\bibitem [{\citenamefont {Narayanan}(2023)}]{narayanan2023understanding}%
  \BibitemOpen
  \bibfield  {author} {\bibinfo {author} {\bibfnamefont {A.}~\bibnamefont {Narayanan}},\ }\href {https://doi.org/10.7916/khdk-m460} {\bibinfo {title} {Understanding social media recommendation algorithms}} (\bibinfo {year} {2023})\BibitemShut {NoStop}%
\bibitem [{\citenamefont {Yang}\ \emph {et~al.}(2018)\citenamefont {Yang}, \citenamefont {Zhu}, \citenamefont {Chen}, \citenamefont {Ma}, \citenamefont {Lv},\ and\ \citenamefont {Lin}}]{yang2018neural}%
  \BibitemOpen
  \bibfield  {author} {\bibinfo {author} {\bibfnamefont {J.}~\bibnamefont {Yang}}, \bibinfo {author} {\bibfnamefont {S.}~\bibnamefont {Zhu}}, \bibinfo {author} {\bibfnamefont {T.}~\bibnamefont {Chen}}, \bibinfo {author} {\bibfnamefont {Y.}~\bibnamefont {Ma}}, \bibinfo {author} {\bibfnamefont {N.}~\bibnamefont {Lv}},\ and\ \bibinfo {author} {\bibfnamefont {J.}~\bibnamefont {Lin}},\ }\bibfield  {title} {\bibinfo {title} {Neural network based min-entropy estimation for random number generators},\ }in\ \href {https://doi.org/10.1007/978-3-030-01704-0_13} {\emph {\bibinfo {booktitle} {International Conference on Security and Privacy in Communication Systems}}}\ (\bibinfo {organization} {Springer},\ \bibinfo {year} {2018})\ pp.\ \bibinfo {pages} {231--250}\BibitemShut {NoStop}%
\bibitem [{\citenamefont {Calude}\ \emph {et~al.}(2010)\citenamefont {Calude}, \citenamefont {Dinneen}, \citenamefont {Dumitrescu},\ and\ \citenamefont {Svozil}}]{calude2010experimental}%
  \BibitemOpen
  \bibfield  {author} {\bibinfo {author} {\bibfnamefont {C.~S.}\ \bibnamefont {Calude}}, \bibinfo {author} {\bibfnamefont {M.~J.}\ \bibnamefont {Dinneen}}, \bibinfo {author} {\bibfnamefont {M.}~\bibnamefont {Dumitrescu}},\ and\ \bibinfo {author} {\bibfnamefont {K.}~\bibnamefont {Svozil}},\ }\bibfield  {title} {\bibinfo {title} {Experimental evidence of quantum randomness incomputability},\ }\href {https://doi.org/10.1103/PhysRevA.82.022102} {\bibfield  {journal} {\bibinfo  {journal} {Physical Review A}\ }\textbf {\bibinfo {volume} {82}},\ \bibinfo {pages} {022102} (\bibinfo {year} {2010})}\BibitemShut {NoStop}%
\bibitem [{\citenamefont {Li}\ \emph {et~al.}(2020)\citenamefont {Li}, \citenamefont {Zhang}, \citenamefont {Sang}, \citenamefont {Gong}, \citenamefont {Wang}, \citenamefont {Wang},\ and\ \citenamefont {Wang}}]{li2020deep}%
  \BibitemOpen
  \bibfield  {author} {\bibinfo {author} {\bibfnamefont {C.}~\bibnamefont {Li}}, \bibinfo {author} {\bibfnamefont {J.}~\bibnamefont {Zhang}}, \bibinfo {author} {\bibfnamefont {L.}~\bibnamefont {Sang}}, \bibinfo {author} {\bibfnamefont {L.}~\bibnamefont {Gong}}, \bibinfo {author} {\bibfnamefont {L.}~\bibnamefont {Wang}}, \bibinfo {author} {\bibfnamefont {A.}~\bibnamefont {Wang}},\ and\ \bibinfo {author} {\bibfnamefont {Y.}~\bibnamefont {Wang}},\ }\bibfield  {title} {\bibinfo {title} {Deep learning-based security verification for a random number generator using white chaos},\ }\href {https://doi.org/10.3390/e22101134} {\bibfield  {journal} {\bibinfo  {journal} {Entropy}\ }\textbf {\bibinfo {volume} {22}},\ \bibinfo {pages} {1134} (\bibinfo {year} {2020})}\BibitemShut {NoStop}%
\bibitem [{\citenamefont {Gutierrez}(2022)}]{gutierrez2022attacking}%
  \BibitemOpen
  \bibfield  {author} {\bibinfo {author} {\bibfnamefont {J.}~\bibnamefont {Gutierrez}},\ }\bibfield  {title} {\bibinfo {title} {Attacking the linear congruential generator on elliptic curves via lattice techniques},\ }\href {https://doi.org/10.1007/s12095-021-00535-6} {\bibfield  {journal} {\bibinfo  {journal} {Cryptography and Communications}\ }\textbf {\bibinfo {volume} {14}},\ \bibinfo {pages} {505} (\bibinfo {year} {2022})}\BibitemShut {NoStop}%
\bibitem [{\citenamefont {Mongia}\ \emph {et~al.}(2024)\citenamefont {Mongia}, \citenamefont {Kumar}, \citenamefont {Prabhakar}, \citenamefont {Banerji},\ and\ \citenamefont {Singh}}]{mongia2024investigating}%
  \BibitemOpen
  \bibfield  {author} {\bibinfo {author} {\bibfnamefont {V.}~\bibnamefont {Mongia}}, \bibinfo {author} {\bibfnamefont {A.}~\bibnamefont {Kumar}}, \bibinfo {author} {\bibfnamefont {S.}~\bibnamefont {Prabhakar}}, \bibinfo {author} {\bibfnamefont {A.}~\bibnamefont {Banerji}},\ and\ \bibinfo {author} {\bibfnamefont {R.~P.}\ \bibnamefont {Singh}},\ }\bibfield  {title} {\bibinfo {title} {Investigating device-independent quantum random number generation},\ }\href {https://doi.org/10.1016/j.physleta.2024.129954} {\bibfield  {journal} {\bibinfo  {journal} {Physics Letters A}\ }\textbf {\bibinfo {volume} {526}},\ \bibinfo {pages} {129954} (\bibinfo {year} {2024})}\BibitemShut {NoStop}%
\bibitem [{\citenamefont {Tsurumaru}\ and\ \citenamefont {Hayashi}(2013)}]{tsurumaru2013dual}%
  \BibitemOpen
  \bibfield  {author} {\bibinfo {author} {\bibfnamefont {T.}~\bibnamefont {Tsurumaru}}\ and\ \bibinfo {author} {\bibfnamefont {M.}~\bibnamefont {Hayashi}},\ }\bibfield  {title} {\bibinfo {title} {Dual universality of hash functions and its applications to quantum cryptography},\ }\href {https://doi.org/10.1109/TIT.2013.2250576} {\bibfield  {journal} {\bibinfo  {journal} {IEEE transactions on information theory}\ }\textbf {\bibinfo {volume} {59}},\ \bibinfo {pages} {4700} (\bibinfo {year} {2013})}\BibitemShut {NoStop}%
\bibitem [{\citenamefont {Zhou}(2020)}]{zhou2020theory}%
  \BibitemOpen
  \bibfield  {author} {\bibinfo {author} {\bibfnamefont {D.-X.}\ \bibnamefont {Zhou}},\ }\bibfield  {title} {\bibinfo {title} {Theory of deep convolutional neural networks: Downsampling},\ }\href {https://doi.org/10.1016/j.neunet.2020.01.018} {\bibfield  {journal} {\bibinfo  {journal} {Neural Networks}\ }\textbf {\bibinfo {volume} {124}},\ \bibinfo {pages} {319} (\bibinfo {year} {2020})}\BibitemShut {NoStop}%
\bibitem [{\citenamefont {Lempel}\ and\ \citenamefont {Ziv}(2003)}]{lempel2003complexity}%
  \BibitemOpen
  \bibfield  {author} {\bibinfo {author} {\bibfnamefont {A.}~\bibnamefont {Lempel}}\ and\ \bibinfo {author} {\bibfnamefont {J.}~\bibnamefont {Ziv}},\ }\bibfield  {title} {\bibinfo {title} {On the complexity of finite sequences},\ }\href {https://doi.org/10.1109/TIT.1976.1055501} {\bibfield  {journal} {\bibinfo  {journal} {IEEE Transactions on information theory}\ }\textbf {\bibinfo {volume} {22}},\ \bibinfo {pages} {75} (\bibinfo {year} {2003})}\BibitemShut {NoStop}%
\bibitem [{\citenamefont {Tsurumaru}\ \emph {et~al.}(2024)\citenamefont {Tsurumaru}, \citenamefont {Ichikawa}, \citenamefont {Takubo}, \citenamefont {Sasaki}, \citenamefont {Lee},\ and\ \citenamefont {Tsutsui}}]{tsurumaru2024indistinguishability}%
  \BibitemOpen
  \bibfield  {author} {\bibinfo {author} {\bibfnamefont {T.}~\bibnamefont {Tsurumaru}}, \bibinfo {author} {\bibfnamefont {T.}~\bibnamefont {Ichikawa}}, \bibinfo {author} {\bibfnamefont {Y.}~\bibnamefont {Takubo}}, \bibinfo {author} {\bibfnamefont {T.}~\bibnamefont {Sasaki}}, \bibinfo {author} {\bibfnamefont {J.}~\bibnamefont {Lee}},\ and\ \bibinfo {author} {\bibfnamefont {I.}~\bibnamefont {Tsutsui}},\ }\bibfield  {title} {\bibinfo {title} {Indistinguishability between quantum randomness and pseudorandomness under efficiently calculable randomness measures},\ }\href {https://doi.org/10.1103/PhysRevA.109.022243} {\bibfield  {journal} {\bibinfo  {journal} {Physical Review A}\ }\textbf {\bibinfo {volume} {109}},\ \bibinfo {pages} {022243} (\bibinfo {year} {2024})}\BibitemShut {NoStop}%
\bibitem [{\citenamefont {Williams}(1991)}]{williams1991extremely}%
  \BibitemOpen
  \bibfield  {author} {\bibinfo {author} {\bibfnamefont {R.~N.}\ \bibnamefont {Williams}},\ }\bibfield  {title} {\bibinfo {title} {An extremely fast {Ziv-Lempel} data compression algorithm},\ }in\ \href {https://doi.org/10.1109/DCC.1991.213344} {\emph {\bibinfo {booktitle} {1991 Data Compression Conference}}}\ (\bibinfo {organization} {IEEE Computer Society},\ \bibinfo {year} {1991})\ pp.\ \bibinfo {pages} {362--363}\BibitemShut {NoStop}%
\bibitem [{\citenamefont {Vadhan}(2012)}]{TCS-010}%
  \BibitemOpen
  \bibfield  {author} {\bibinfo {author} {\bibfnamefont {S.~P.}\ \bibnamefont {Vadhan}},\ }\bibfield  {title} {\bibinfo {title} {Pseudorandomness},\ }\href {https://doi.org/10.1561/0400000010} {\bibfield  {journal} {\bibinfo  {journal} {Foundations and Trends® in Theoretical Computer Science}\ }\textbf {\bibinfo {volume} {7}},\ \bibinfo {pages} {1} (\bibinfo {year} {2012})}\BibitemShut {NoStop}%
\bibitem [{\citenamefont {Matsumoto}\ and\ \citenamefont {Nishimura}(1998)}]{matsumoto1998mersenne}%
  \BibitemOpen
  \bibfield  {author} {\bibinfo {author} {\bibfnamefont {M.}~\bibnamefont {Matsumoto}}\ and\ \bibinfo {author} {\bibfnamefont {T.}~\bibnamefont {Nishimura}},\ }\bibfield  {title} {\bibinfo {title} {Mersenne twister: a 623-dimensionally equidistributed uniform pseudo-random number generator},\ }\href {https://doi.org/10.1145/272991.272995} {\bibfield  {journal} {\bibinfo  {journal} {ACM Transactions on Modeling and Computer Simulation (TOMACS)}\ }\textbf {\bibinfo {volume} {8}},\ \bibinfo {pages} {3} (\bibinfo {year} {1998})}\BibitemShut {NoStop}%
\bibitem [{\citenamefont {Bernstein}\ and\ \citenamefont {Vazirani}(1993)}]{Vazirani_1993_QCT}%
  \BibitemOpen
  \bibfield  {author} {\bibinfo {author} {\bibfnamefont {E.}~\bibnamefont {Bernstein}}\ and\ \bibinfo {author} {\bibfnamefont {U.}~\bibnamefont {Vazirani}},\ }\bibfield  {title} {\bibinfo {title} {Quantum complexity theory},\ }in\ \href {https://doi.org/10.1145/167088.167097} {\emph {\bibinfo {booktitle} {Proceedings of the Twenty-Fifth Annual ACM Symposium on Theory of Computing}}}\ (\bibinfo  {publisher} {Association for Computing Machinery},\ \bibinfo {address} {New York, NY, USA},\ \bibinfo {year} {1993})\ p.\ \bibinfo {pages} {11–20}\BibitemShut {NoStop}%
\end{thebibliography}%

\end{document}